\documentclass[sigconf,screen]{acmart}

\usepackage{xurl}
\usepackage{xspace}
\usepackage{cancel}
\usepackage{pifont}
\usepackage{comment}
\usepackage{listings}
\usepackage{multirow}
\usepackage{booktabs}
\usepackage[boxed,lined,noend]{algorithm2e}
\usepackage[dvipsnames]{xcolor}
\usepackage{enumitem}
\newcommand{\para}[1]{\vspace{.05in} \noindent \textbf{#1}}
\usepackage{fancyhdr}
\usepackage[normalem]{ulem}
\usepackage{flushend}
\usepackage{soul}
\usepackage{tikz}
\usepackage{cleveref}
\usepackage{graphicx}
\usepackage{upquote}
\usepackage{xcolor,beramono}
\usepackage{microtype}
\usepackage[normalem]{ulem}
\usepackage{subcaption}
\definecolor{yellow}{rgb}{1.00, 1.00, 0.0}

\newcommand*\highlightht{2.8ex}
\newcommand*\highlightdp{-.8ex}
\newcommand*\highlightwd{0.2ex}

\newcommand\highlightcommon[1]
  {%
    \bgroup
    \markoverwith
      {\textcolor{#1}{\smash{\rule[\highlightdp]{\highlightwd}{\highlightht}}}}%
    \ULon
  }

\newcommand*\circled[1]{\tikz[baseline=(char.base)]{
            \node[shape=circle,draw,inner sep=1pt] (char) {#1};}}

\makeatletter
\newcommand{\crefnames}[3]{%
  \@for\next:=#1\do{%
    \expandafter\crefname\expandafter{\next}{#2}{#3}%
  }%
}
\makeatother
\crefnames{part,chapter,section}{\S}{\S\S}
\crefnames{para,paragraph,subparagraph}{\P}{\P\P}

\SetAlFnt{\small}
\NoCaptionOfAlgo
\SetKwIF{If}{ElseIf}{Else}{if}{}{else if}{else}{end if}%
\makeatletter
\newcommand{\removelatexerror}{\let\@latex@error\@gobble}
\makeatother

\newcommand{\OmitForASPLOS}[1]{}

\makeatletter
\newenvironment{btHighlight}[1][]
{\begingroup\tikzset{bt@Highlight@par/.style={#1}}\begin{lrbox}{\@tempboxa}}
{\end{lrbox}\bt@HL@box[bt@Highlight@par]{\@tempboxa}\endgroup}

\newcommand\btHL[1][]{%
  \begin{btHighlight}[#1]\bgroup\aftergroup\bt@HL@endenv%
}
\def\bt@HL@endenv{%
  \end{btHighlight}%
  \egroup
}
\newcommand{\bt@HL@box}[2][]{%
  \tikz[#1]{%
    \pgfpathrectangle{\pgfpoint{1pt}{0pt}}{\pgfpoint{\wd #2}{\ht #2}}%
    \pgfusepath{use as bounding box}%
    \node[anchor=base west, fill=orange!30,outer sep=0pt,inner xsep=1pt, inner ysep=0pt, rounded corners=3pt, minimum height=\ht\strutbox+1pt,#1]{\raisebox{1pt}{\strut}\strut\usebox{#2}};
  }%
}
\makeatother

\AtBeginDocument{%
  \providecommand\BibTeX{{%
    \normalfont B\kern-0.5em{\scshape i\kern-0.25em b}\kern-0.8em\TeX}}}

\setcopyright{acmcopyright}
\acmPrice{15.00}
\acmDOI{10.1145/3445814.3446708}
\acmYear{2021}
\copyrightyear{2021}
\acmSubmissionID{asplos21main-p148-p}
\acmISBN{978-1-4503-8317-2/21/04}
\acmConference[ASPLOS '21]{Proceedings of the 26th ACM International Conference on Architectural Support for Programming Languages and Operating Systems}{April 19--23, 2021}{Virtual, USA}
\acmBooktitle{Proceedings of the 26th ACM International Conference on Architectural Support for Programming Languages and Operating Systems (ASPLOS '21), April 19--23, 2021, Virtual, USA}

\newcounter{BalanceAtReference}
\newcounter{ReferenceIndexForBalancing}

\makeatletter

\global\@ACM@balancefalse

\def\@balancelastpageonce{%
  \ifnum\value{ReferenceIndexForBalancing}=\value{BalanceAtReference}
    \newpage
  \else
    \relax
  \fi
  \stepcounter{ReferenceIndexForBalancing}
}
\pretocmd{\bibitem}{\@balancelastpageonce}
  {} 
  {\@latex@error{Patching \bibitem failed}{\@ehd}}

\makeatother

\begin{document}

\title[Speculative Interference Attacks: Breaking Invisible Speculation Schemes]{Speculative Interference Attacks: \\ Breaking Invisible Speculation Schemes}


\author{Mohammad Behnia}
\affiliation{%
  \institution{UIUC, USA}
  \streetaddress{}
  \city{}
  \state{}
  \country{}
  \postcode{}
}

\author{Prateek Sahu}
\affiliation{%
  \institution{UT Austin, USA}
  \streetaddress{}
  \city{}
  \state{}
  \country{}
  \postcode{}}

\author{Riccardo Paccagnella}
\affiliation{%
  \institution{UIUC, USA}
  \streetaddress{}
  \city{}
  \state{}
  \country{}
  \postcode{}}

\author{Jiyong Yu}
\affiliation{%
  \institution{UIUC, USA}
  \streetaddress{}
  \city{}
  \state{}
  \country{}
  \postcode{}}

\author{Zirui Neil Zhao}
\affiliation{%
  \institution{UIUC, USA}
  \streetaddress{}
  \city{}
  \state{}
  \country{}
  \postcode{}}

\author{Xiang Zou}
\affiliation{%
  \institution{Intel Corporation, USA}
  \streetaddress{}
  \city{}
  \state{}
  \country{}
  \postcode{}}

\author{Thomas Unterluggauer}
\affiliation{%
  \institution{Intel Corporation, USA}
  \streetaddress{}
  \city{}
  \state{}
  \country{}
  \postcode{}}

\author{Josep Torrellas}
\affiliation{%
  \institution{UIUC, USA}
  \streetaddress{}
  \city{}
  \state{}
  \country{}
  \postcode{}}

\author{Carlos Rozas}
\affiliation{%
  \institution{Intel Corporation, USA}
  \streetaddress{}
  \city{}
  \state{}
  \country{}
  \postcode{}}

\author{Adam Morrison}
\affiliation{%
  \institution{Tel Aviv University, Israel}
  \streetaddress{}
  \city{}
  \state{}
  \country{}
  \postcode{}}

\author{Frank Mckeen}
\affiliation{%
  \institution{Intel Corporation, USA}
  \streetaddress{}
  \city{}
  \state{}
  \country{}
  \postcode{}}

\author{Fangfei Liu}
\affiliation{%
  \institution{Intel Corporation, USA}
  \streetaddress{}
  \city{}
  \state{}
  \country{}
  \postcode{}}

\author{Ron Gabor}
\affiliation{%
  \institution{Toga Networks, Israel}
  \streetaddress{}
  \city{}
  \state{}
  \country{}
  \postcode{}}

\author{Christopher W. Fletcher}
\affiliation{%
  \institution{UIUC, USA}
  \streetaddress{}
  \city{}
  \state{}
  \country{}
  \postcode{}}

\author{Abhishek Basak}
\affiliation{%
  \institution{Intel Corporation, USA}
  \streetaddress{}
  \city{}
  \state{}
  \country{}
  \postcode{}}

\author{Alaa Alameldeen}
\affiliation{%
  \institution{Simon Fraser University, Canada}
  \streetaddress{}
  \city{}
  \state{}
  \country{}
  \postcode{}}

\renewcommand{\shortauthors}{Behnia et al.}

\begin{abstract}
    Recent security vulnerabilities that target speculative execution (e.g., Spectre) present a significant challenge for processor design. 
    These highly publicized vulnerabilities use
    speculative execution to learn victim secrets by changing the cache state. As a result, recent computer architecture research has focused on \emph{invisible speculation} mechanisms that attempt to block changes in cache state due to speculative execution. Prior work has shown significant success in preventing Spectre and other attacks at modest performance costs.
    
    In this paper, we introduce \emph{speculative interference attacks}, which show that prior invisible speculation mechanisms do not fully block speculation-based attacks that use cache state. We make two key observations. First, \emph{mis-speculated younger instructions can change the timing of older, bound-to-retire instructions,} including memory operations. Second, changing the timing of a memory operation can \emph{change the order of that memory operation relative to other memory operations,} resulting in persistent changes to the cache state. Using both of these observations, we demonstrate (among other attack variants) that secret information accessed by mis-speculated instructions can change the order of bound-to-retire loads. Load timing changes can therefore leave secret-dependent changes in the cache, even in the presence of invisible speculation mechanisms.
    
    We show that this problem is not easy to fix. Speculative interference converts timing changes to persistent cache-state changes, and timing is typically ignored by many cache-based defenses. We develop a framework to understand the attack and demonstrate concrete proof-of-concept attacks against invisible speculation mechanisms. 
    We conclude with a discussion of 
    security definitions that are sufficient to block the attacks, along with preliminary defense ideas based on those definitions. 
    \newline
\end{abstract}

\ccsdesc[500]{Security and privacy~Side-channel analysis and countermeasures}

\keywords{invisible speculation, speculative execution attacks, microarchitectural covert channels}

\maketitle

\newcommand{\ISPEC}{\cite{safespec,invisispec,sakalis_isca19,conditional_spec,muontrap}\xspace}


\section{Introduction}
\label{sec:introduction}

\newcommand{\accessaddr}{\texttt{addr}}
\newcommand{\secret}{{\color{ForestGreen}\texttt{secret}}\xspace}

Speculative execution attacks such as Spectre~\cite{spectre} and follow-on work~\cite{netspectre,speculative_buffer_overflow,invisispec,spec_variant_four,sgx_spectre,ret2spec,spectre_returns,smother} have opened a new chapter in processor security. 
In these attacks, adversary-controlled transient instructions---i.e., speculative instructions bound to squash---access and then transmit potentially sensitive program data over \emph{microarchitectural covert channels} (e.g., the cache~\cite{flush+}, port contention~\cite{smother}).
For example in Spectre variant 1---\texttt{if (i < N) \{ j = A[i]; B[j]; \}}---speculative execution bypasses a bounds check due to a branch misprediction, accesses an out-of-bounds value (\texttt{j = A[i]}) and transmits that value through a \emph{cache-based covert channel} (\texttt{B[j]}), i.e., by forcing a cache fill to occur in a set that depends on \texttt{j}.
In this paper, we consider the illegally accessed value \texttt{j} to be the \emph{secret}.
Here, the attacker controls the value of \texttt{i}, thus \texttt{j} can be any value in program memory and the covert channel can reveal arbitrary program data.

While a variety of covert channels can be used to leak secret values under mis-speculation, cache-based covert channels~\cite{spectre,cache_sc,flush+,cache_bleed,noninclusive,last_level_cache_practical} make the fewest assumptions on the attacker and have therefore received the most attention.
This is for two reasons.
First, secret-dependent cache fills leave a persistent footprint in the cache which is observable long after speculation squashes.
Second,
certain levels of modern cache hierarchies
are globally shared by all cores in the system, enabling attackers to observe said persistent state changes from other physical cores~\cite{last_level_cache_practical,noninclusive}.
By contrast, 
many other covert channels (e.g., arithmetic port contention~\cite{portsmash,smother}) 
leave only intermittent side effects that must be monitored before the squash, and/or require that the attacker share hardware resources on the same physical core (e.g., 
branch predictor channels~\cite{BranchScope,evtyushkin2016understanding,branchpred_sc,evtyushkin2016jump})---both of which can be easily blocked (e.g., disabling SMT).

The above view of the covert channel landscape has led to a surge of architecture-level ``\emph{invisible speculation}'' proposals 
to block cache-based covert channels due to mis-speculations 
(e.g., InvisiSpec~\cite{invisispec}, SafeSpec~\cite{safespec}, Delay-on-Miss~\cite{sakalis_isca19}, Conditional Speculation~\cite{conditional_spec}, 
MuonTrap~\cite{muontrap}).
Invisible speculation schemes add hardware to
prevent mis-speculated loads from making persistent state changes to the memory subsystem.
To maintain the performance benefits of caching, only non-speculative loads that are bound to retire are allowed to modify the cache state.
To maintain the performance benefits of out-of-order execution, loads are allowed to ``invisibly'' execute (i.e., bring data directly to the core without filling the cache) and forward their results to dependent instructions.

\para{This Paper.} In this paper we introduce \emph{speculative interference attacks}, which show that invisible speculation schemes do not fully block speculation-based attacks that use the cache state.
Our attacks are based on two key observations.
First, that mis-speculated instructions can influence the timing of older, bound-to-retire operations.  
Second, 
if changing the timing of a memory operation changes the \emph{order} of that memory operation \emph{with respect to other memory operations}, the resulting reordering can cause persistent cache-state changes.
Putting these together, we show (among other attack variants) how secret information accessed in a mis-speculated window influences the order of bound-to-retire loads, leaving secret-dependent state changes in the cache---even if invisible speculation is enabled. 

To explain these ideas, consider a simple but representative invisible speculation scheme: \emph{Delay-on-miss} (DoM)~\cite{sakalis_isca19}.
DoM issues a speculative load and 
(a) on an L1 cache hit, forwards the load result to dependent instructions, or 
(b) on an L1 cache miss, delays servicing the miss and re-issues the load when it becomes non-speculative.
In case (a), DoM does not update any replacement state (e.g., replacement bits) in the L1 cache until the load becomes non-speculative.
For simplicity, we explain ideas assuming only branch instructions cast speculative shadows~\cite{sakalis_isca19}, i.e., a load is considered non-speculative/safe iff it is older than the oldest unresolved branch.
We discuss attacks on more conservative DoM variants in \cref{sec:break}.

DoM's (and other invisible speculation schemes') stated security goal is to only focus on blocking cache state changes due to mis-speculations, while leaving other covert channels un-blocked.
This is reflected in DoM's design.
On the one hand, DoM prevents mis-speculated loads from directly changing the cache state.
On the other hand, DoM allows mis-speculated loads to forward their results to dependent instructions, which can clearly form covert channels through intermittent state changes.
In fact, this is exactly the basis for forming, e.g., arithmetic unit port contention covert channels~\cite{smother,portsmash}.

\begin{figure}[h]
  \centering\includegraphics[width=\columnwidth]{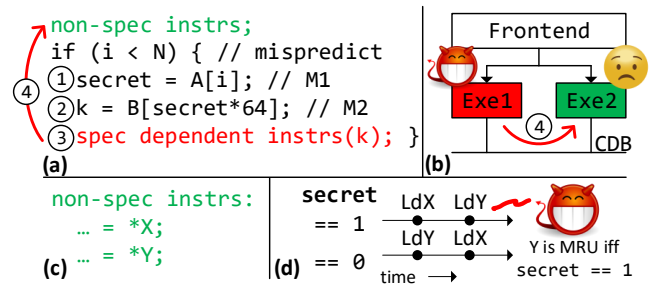}
\caption{\small \textbf{
Speculative interference example.
\texttt{secret} is 0 or 1.
(a) Assume the code snippet is run on a processor protected by invisible speculation such as DoM and that \texttt{\&B[0]} is cached while \texttt{\&B[64]} is not cached.
(b) This results in \emph{speculative dependent instructions} conditionally contending for execution resources with \emph{non-speculative instructions}, depending on the value of \texttt{secret}.
(c), (d) If the non-speculative instructions are two loads, the contention can influence the order in which the loads are issued.
Finally, the attacker can infer the secret based on the cache replacement state after the loads both issue.}}

\label{fig:spec_interference}
\end{figure}

This paper demonstrates how instructions that cause intermittent state changes can be leveraged to create persistent state changes in the cache.
Consider the example in Figure~\ref{fig:spec_interference}, modeled after Spectre variant 1.
Suppose this code is run on a processor using DoM.
In Figure~\ref{fig:spec_interference} (a), a mis-speculated load M1 forwards secret data \texttt{secret} to a second load M2 (\ding{192}).
A normal Spectre attack would monitor the cache state change left by M2 to deduce \texttt{secret}.
To prevent this leak, DoM would prevent M2 from changing the cache state, specifically by allowing it to access and return data from the L1 if there is an L1 hit and delaying its execution otherwise.
While this blocks the cache state change due to M2, M2 is allowed to forward its result when it completes (\ding{193}), meaning that dependent instructions execute at a time that depends on \texttt{secret} (\ding{194}).
This has the potential to create a traditional non-cache based covert channel, e.g., through execution unit port usage, which DoM ignores.

Our key observation is that secret-dependent timing effects caused by the dependent instructions can be monitored \emph{indirectly} through how they interact with the execution of \emph{older non-speculative instructions} (\ding{195}).
Although the non-speculative instructions come before the speculative dependent instructions in program order, out-of-order execution could have both of them executing concurrently and contending for resources.
In our example (Figure~\ref{fig:spec_interference} (b)), the non-speculative and speculative dependent instructions use execution units EXE1 and EXE2, respectively, and contend for the common data bus (CDB) in the same cycle.
We call this \emph{speculative interference}.

Next, we show how speculative interference can be used to bootstrap a change in the cache state.
Specifically, suppose the non-speculative instructions are made up of two independent loads to addresses X and Y in different cache lines mapped to the same set, shown in Figure~\ref{fig:spec_interference} (c).
Since these loads are older than the mispredicted branch in program order, they are not protected by DoM.
We show how, depending on the timing changes caused by the speculative dependent instructions, the order in which the load to X is issued with respect to the load to Y can change.
\emph{That is, depending on a secret, the processor issues either loads to X followed by Y or Y followed by X.} 
To finish the attack (Figure~\ref{fig:spec_interference} (d)), we show how changing the order of memory operations can be used to create persistent changes in the cache state, 
the intuition being that state in the cache (e.g., replacement bits) depends on not just what requests are made, but also their order.

This issue is not easy to fix.
The crux of the problem is that timing changes can be converted to persistent state changes.
These timing changes can arise due to interference through a large number of microarchitectural structures, through different instructions, etc. 
Further, while our example reorders two loads that originate from the same thread, there are many other memory address streams through which to interleave operations, e.g., interleaving instruction and data cache accesses, accesses made across threads and security domains, etc.---which further widens the attack surface.
The rest of the paper expands on the above ideas as follows:
\begin{itemize}[nosep,leftmargin=1em,labelwidth=*,align=left]
    \item 
    We introduce and provide a framework to reason about \emph{speculative interference attacks}, whereby subtle secret-dependent microarchitectural interference influences the behavior of older non-speculative instructions.
    We show how 
    this can be used to create cache-based covert channels, even in the presence of invisible speculation schemes.
    \item We implement proof-of-concept exploits for three 
    such attacks on a real machine, exploiting interference in the processor's out-of-order issue logic, MSHR usage and frontend queues.

    All attacks are cache-based and work across physical cores.

    \item 
    We provide a sufficiently strong security definition to block the attacks, and also provide a starting point defense and discussion to set a research agenda for more efficiently and comprehensively addressing the problem.
\end{itemize}

\section{Background} \label{sec:background}

\subsection{Threat Model} \label{sec:threatmodel}

We adopt the standard threat model used by invisible speculation schemes~\ISPEC.
Such schemes care about preventing ``persistent'' side effects that are observable due to mis-speculated loads.
For example, which lines are in the cache, replacement and coherence state of each line, etc.

Invisible speculation schemes further distinguish from where the attacker is monitoring the covert channel (i.e., where the receiver~\cite{dawg} runs).
In particular, one of the first invisible speculation schemes by Yan et al.~\cite{invisispec} specifies several such attacker models:

\textbf{SameThread model}:
Here we consider untrusted code interleaved with trusted code, as in the case of a sandbox.

\textbf{CrossCore model}:
Here,
the idea is that the system prevents untrusted code from running on a sibling hyper-thread. However, the receiver may run on another core and monitor a cache-based covert channel through a shared cache.

We will show attacks against these models.
Yan et al.~\cite{invisispec} also specifies an \textbf{SMT} model where the receiver runs in an adjacent hyper-thread.
This gives the attacker more power, thus we will focus on the former two models.

\subsection{OoO Processor Pipeline}

Dynamically scheduled processors execute instructions out of order (OoO) to improve performance~\cite{tomasulo1967efficient,hennessy2011computer}.
An instruction is \emph{fetched} by the processor frontend, {\em dispatched} to reservation stations (RS) for
scheduling, {\em issued} to execution units (EUs) in the processor backend, and finally {\em retired} when it updates the machine's architected state.
Instructions proceed through the frontend, backend and retirement stages in order, possibly out of order and in order, respectively.
In-order retirement is implemented by queueing instructions in a reorder buffer
(ROB)~\cite{johnson1991superscalar} in program order and retiring a completed instruction when it reaches the ROB
head.

\subsection{Invisible Speculation Schemes}
\label{background:invisi_specs}

Invisible speculation aims to block covert channels from forming through the cache due to mis-speculated loads~\ISPEC.
While speculative execution attacks can leverage both cache-based and non-cache-based covert channels, invisible speculation schemes are only concerned with attacks using the cache because cache-state changes are relatively simple to monitor.
Specifically, cache-state changes \emph{persist} after the squash and can be measured across physical cores (as in the CrossCore model).
By contrast, non-persistent covert channels, such as those through execution units~\cite{Mult_leaky} and ports~\cite{smother,gras2020absynthe}, are more difficult to monitor and place additional restrictions on the attacker (e.g., limit the attacker to the SMT model~\cite{smother}).

\begin{figure*}[t]
\centering
\includegraphics[width=\textwidth]{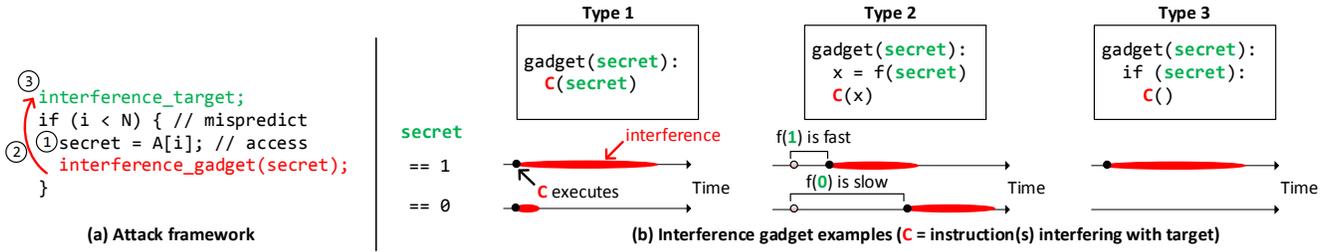}
\caption{\small \textbf{ Speculative interference attack framework (a) and classification of interference gadgets (b).
}
}
\label{fig:framework}
\end{figure*}

While there have been multiple invisible speculation proposals, they all share several common traits.
For security, they prevent state changes to the cache due to mis-speculated loads, until those loads are deemed safe/become non-speculative. 
Performance hinges on several optimizations.
First, speculative loads should---subject to the constraint of not updating cache state---be allowed to issue and forward their results to dependent (speculative) instructions.
This allows these schemes to maintain efficiency in the common case that speculation turns out to be correct.
Second, loads should be able to update cache state when they become non-speculative (safe).
Together, these performance optimizations enable such schemes to reap the benefits of out-of-order execution and caching.

Different invisible speculation schemes differ in their exact policies for allowing speculative loads to issue.
We describe the ``Delay-On-Miss'' (DOM)~\cite{sakalis_isca19} scheme here,
as it is simple and illustrates the main ideas.

\para{Delay-On-miss (DOM)}
allows loads that hit in the L1 cache to execute and forward their results to dependent instructions (which are themselves allowed to execute).
Any cache-state change that would have been made as a by-product of the L1 cache hit (e.g., modifying replacement bits) is deferred until the load becomes non-speculative.
If the load misses the L1 cache, it is delayed and re-executed when it becomes non-speculative.

\section{Speculative Interference Attacks} \label{sec:break}

We now describe \emph{speculative interference attacks}. These attacks exploit the fact
that while invisible speculation blocks direct cache-state changes by issued speculative loads, it does not restrict how secrets returned by those loads flow through the pipeline and impact the execution of non-memory instructions (\cref{background:invisi_specs}).
Our key insight is that the secret-dependent resource usage patterns of such non-memory speculative instructions \emph{can be transformed into} cache-based covert channels.
Specifically, a speculative interference attack exploits intermittent resource contention to influence the timing of \emph{non-speculative}
parts of the pipeline (\cref{sec:framework}).  These timing effects are then used
to make the non-speculative cache access pattern depend on the secret, creating a cache covert channel (\cref{sec:exploit}).

\subsection{Making Mis-speculation Influence the Timing of Non-speculative Actions} \label{sec:framework}

We first present a framework for leaking a secret bit by making the secret determine the time at which an unprotected
\emph{victim} memory operation accesses the cache, modifying its state.
See Figure~\ref{fig:framework}(a) for a visualization.
\circled{1}
A mis-speculated \emph{access} load reads a secret into the pipeline and forwards that secret to a sequence of mis-speculated instructions, called the \emph{interference gadget}, which
creates secret-dependent pressure on some microarchitectural resource(s).
\circled{2} Contention on these resource(s) influences the timing of actions performed by a non-speculative part of the pipeline, called the
\emph{interference target}, making them encode the secret. \circled{3} The target is chosen
so that changing its timing creates a pipeline ``ripple effect'' that ultimately delays an unprotected victim memory access.
That is, how quickly the target executes determines when the unprotected victim's memory operation is issued and
accesses the cache, thereby modifying the cache state.

Interference gadgets can be classified as one of three types, determined by how the secret is
used to create resource contention interfering with the target's execution. Figure~\ref{fig:framework}(b)
shows examples of the gadget types.

A \emph{Type 1} gadget forwards the secret to instruction(s) with operand-dependent resource usage patterns, called
\emph{transmitters}~\cite{stt}, which issue at a secret-independent time.
The resulting secret-dependent resource pressure
during the transmitter's execute stage
interferes with the target.
Examples of transmitters include data-dependent arithmetic~\cite{Mult_leaky} or loads.
Notice that a (speculative) load is considered a transmitter despite not modifying the cache state under invisible speculation. The reason is that
its usage pattern of other resources, such as MSHRs, depends on its address operand. 

In a \emph{Type 2} gadget, the secret is encoded through instructions issuing at a secret-dependent time.
(As opposed to a Type 1 gadget, in which the interfering instructions issue at secret-independent times but have secret-dependent resource usage during their execute stage.)
Secret-dependent issue time is achieved by having the interfering instructions be data-dependent on variable-latency instruction(s) that
data-depend on the secret. Importantly, while these variable-latency instruction(s) are necessarily transmitter(s), here their secret-dependent resource usage does not interfere with the target. It is only used to influence \emph{when} subsequent interfering instructions
(which can be transmitters or non-transmitters) are issued for execution.

In a \emph{Type 3} gadget, whether interfering instructions execute at all depends on the secret.
For example, if the gadget contains a branch with a secret-dependent predicate, then resolution of the branch ``poisons'' the
branch predictor's state with secret-dependent information. As a result, the branch's prediction can become secret-dependent
and determine whether the contending instructions execute in subsequent executions of the gadget.

All gadgets exploit the fact that invisible speculation does not ``protect'' a transmitter's resource usage pattern,
execution time, or branch prediction. Although the resulting secret-dependent execution behavior cannot be directly observed by
the attacker (which monitors the cache), it can be indirectly observed through its influence on the non-speculative interference target,
as discussed next. 

\newcommand{\transmitternote}{Transmitters in the interference gadget are {\btHL[fill=white!10,draw=red,solid,thin]boxed} in both the pseudo-code and timeline to show correspondence.}
\begin{figure*}[t]
    \vspace{0.5ex}
    \centering
    \begin{minipage}{0.35\textwidth}
        \centering
        \begin{lstlisting}[basicstyle=\footnotesize,xleftmargin=0.3cm,escapechar=\%]
A = ...  // takes Z cycles
y = load(A)  // Interference Target
if (i < N): // mispred. taken (miss on N)
   secret = load(&TargetArray[i]) // access
   // Interference Gadget
    x0   = `load(&S[secret * 64 * 0])`
    x1   = `load(&S[secret * 64 * 1])`
    ...
    x$_{M-1}$ = `load(&S[secret * 64 * (M-1)])`
        \end{lstlisting}
        \subcaption{(a)}
    \end{minipage}
    \begin{minipage}{0.58\textwidth}
        \centering
        \includegraphics[trim={0 0.6cm 0 0},clip,width=1\textwidth]{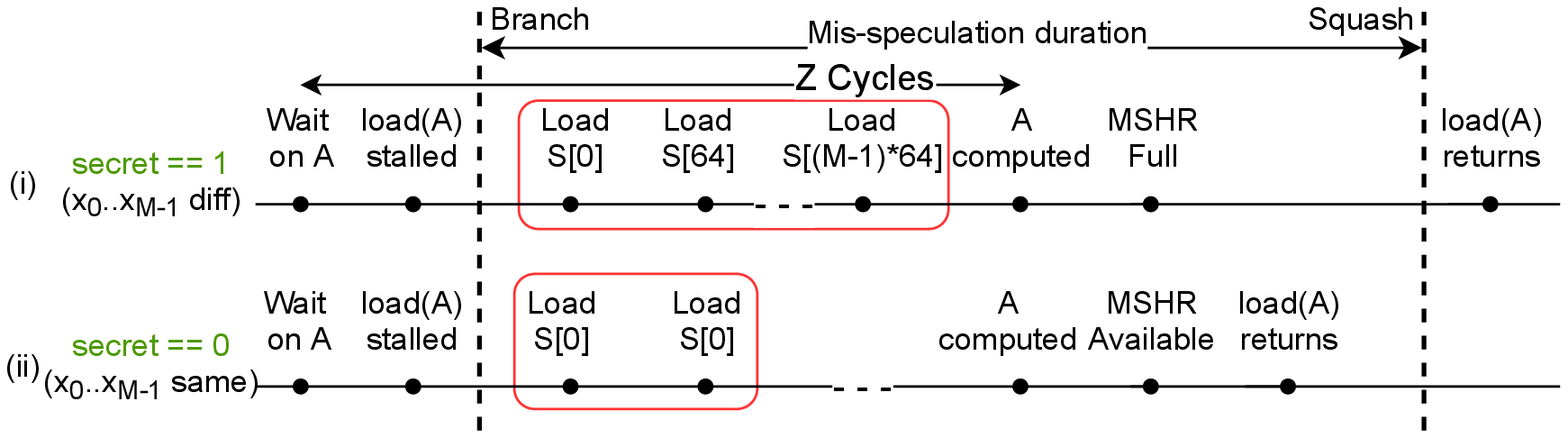} 
        \subcaption{(b)}
    \end{minipage}\hfill
    \vspace{-1ex}
    \caption{\small \textbf{ Delaying a load using miss status holding registers
    ($\mathbf{G^D_{MSHR}}$). $M$ is the number of L1 D-Cache MSHRs.
    \transmitternote}
    }
    \label{fig:lfb}
\end{figure*}

\begin{figure*}[t]
    \centering
    \begin{minipage}{0.35\textwidth}
        \centering
        \begin{lstlisting}[basicstyle=\footnotesize,xleftmargin=0.3cm,escapechar=\%]
z = ...   // takes Z cycles
A = f(z)
y = load(A)  // Interference Target
if (i < N): // mispred. taken (miss on N)
    secret = load(&TargetArray[i]) // access
    // Interference Gadget
    x = `load(&S[secret * 64])`
    f$'$(x)
    \end{lstlisting}
    \subcaption{(a)}
    \end{minipage}
    \begin{minipage}{0.58\textwidth}
        \centering
        \includegraphics[trim={0 0.5cm 0 0},clip,width=1\textwidth]{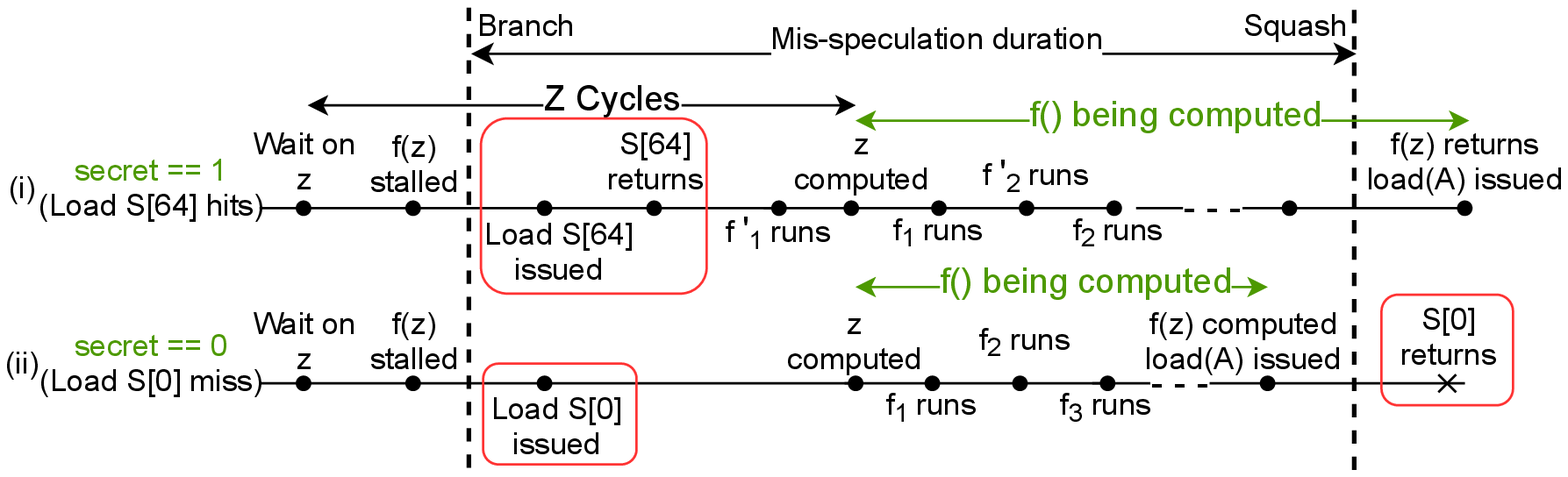}
        \subcaption{(b)}
    \end{minipage}\hfill
    \caption{\small \textbf{ 
    Delaying a load using contention on a non-pipelined EU ($\mathbf{G^D_{NPEU}}$).
Instruction sequences \texttt{f} and \texttt{f'} use the same non-pipelined EU.
\transmitternote
}
}
    \label{fig:dcu}
\end{figure*}

\subsubsection{Interference Gadgets \& Targets} \label{sec:gadgets}

Here, we design several interference gadget/targets, which illustrate Type 1 and 2 gadgets that delay D- and I-Cache accesses.
\cref{sec:attacks} implements the attack variants described in this section and shows that they lead to observable interference in practice.
Our goal here is not to exhaustively enumerate all possible gadgets/targets, but to illustrate the problem. Exploring if (and which) other microarchitectural resources can be used to build interference gadgets is future work.

We first present two gadgets that delay an unprotected D-Cache access.  We exploit the fact that in all invisible speculation
designs,
a load that executes only when it reaches the head of the ROB performs its D-Cache access unprotected.
Our gadgets therefore require interference targets in which a victim load's address, or \emph{target address}, becomes available just as it reaches the head of the ROB.
Our interference gadgets either delay the victim load from reaching the ROB head or delay its execution (cache access)
after it has reached the ROB head.

Next, we present a gadget that delays an unprotected I-Cache access. Such accesses are performed
by InvisiSpec and DoM.  We acknowledge that an unprotected I-Cache access can form a cache-based covert
channel in and of itself but describe this gadget due to its interesting property of interfering with
the frontend and not with some instruction's execution.

\para{$\mathbf{G^D_{MSHR}}$: Delay \underline{d}ata access with \underline{MSHR} contention.}\label{para:mshr}
This is a
Type 1 gadget that delays the execution time of a load at the head of the ROB by a secret-dependent amount of time.
Figure~\ref{fig:lfb}(a) shows the gadget and target.  The target consists of the victim load whose address operand, \texttt{A}, takes
$Z$ cycles to generate.  
The value
$Z$ is such that the gadget's instructions can issue while the target address is
being generated.  The gadget consists of $M$ independent loads, where $M$ is the number of L1 D-Cache miss status handling registers
(MSHRs), each of which holds information on all the outstanding misses for some cache line.
(Here and elsewhere, the gadget executes in the shadow of a slow-to-resolve mispredicted branch, due to a cache miss on \texttt{N}.)
The gadget's goal is to create secret-dependent MSHR pressure, to delay when the victim load obtains its data after reaching the
head of the ROB and issuing.  This gadget targets invisible speculation designs that issue speculative L1 D-Cache misses, i.e., InvisiSpec,
SafeSpec, and MuonTrap.  None of these designs specify changes to the MSHR allocation policy, so we assume they use the standard
policy of allocating an MSHR to a missing load based on issue order.
Figure~\ref{fig:lfb}(b) shows the attack timeline.

\begin{figure*}[t]
    \centering
    \begin{minipage}{0.35\textwidth}
        \centering
        \begin{lstlisting}[basicstyle=\footnotesize,xleftmargin=0.3cm,escapechar=|]
if (i < N): // mispredict taken (miss on N) |\label{line:cond}|
    secret = load(&TargetArray[i])  // access
    // Interference Gadget
    x = `load(&S[secret * 64])` |\label{line:load}|
    // Congest RS
    sum += x;                   |\label{line:alu}|
    ...
    sum += x; // many times
target instr. // Target instruction  |\label{line:tgti}|
\end{lstlisting}
\subcaption{(a)}
    \end{minipage}\hfill
    \begin{minipage}{0.58\textwidth}
        \centering
        \includegraphics[trim={0 0.2cm 0 0},clip,width=1\textwidth]{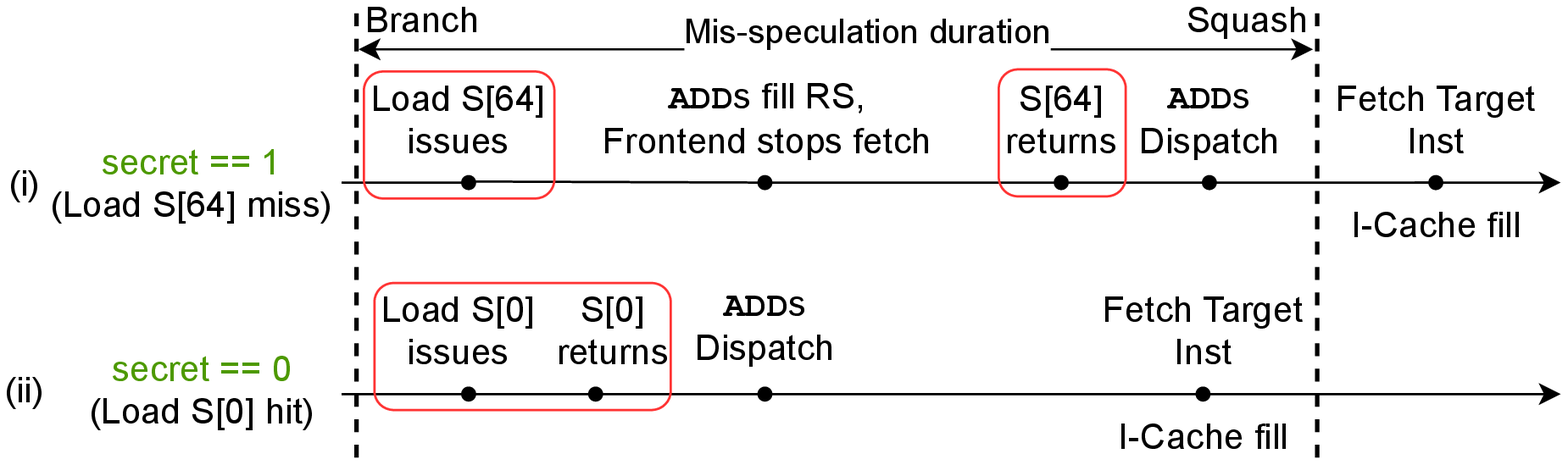}
        \subcaption{(b)}
    \end{minipage}\hfill
    \caption{\small \textbf{ Back-throttling the Fetch Unit by contending for RS for
    $\mathbf{G^I_{RS}}$ gadget. \texttt{sum+=x} repeats for N (number of RS slots) times.
    \transmitternote
}
}
    \label{fig:icu}
\end{figure*}

\noindent \circled{1} If $\mathtt{secret}=1$, each gadget load accesses a different cache line.  The attacker primes the cache so that each
of these accesses is an L1 D-Cache miss.  The result is that each load allocates a distinct MSHR, exhausting the available MSHRs.
Thus, the victim load (assumed to be an L1 D-Cache miss) cannot issue and is delayed until one of the gadget loads
completes or the mis-speculation is squashed.

\noindent \circled{2} If $\mathtt{secret}=0$, all the gadget loads access the same cache line.  They therefore use the same MSHR, which leaves
MSHRs available for the victim load once it reaches the head of the ROB.  The victim load can issue and
is not delayed.

\para{$\mathbf{G^D_{NPEU}}$: Delay \underline{d}ata access using \underline{n}on-\underline{p}ipelined \underline{EU}.} \label{para:dcu}
This is a Type 2 gadget that creates a secret-dependent delay of the victim's target address generation.
Figure~\ref{fig:dcu}(a) shows the gadget and target.  The target consists of a victim load whose address operand, \texttt{A},
is generated by a dependent chain of instructions, denoted \texttt{f}.
The gadget consists of a load (transmitter) and a sequence of
independent instructions, denoted \texttt{f'}, that depend on the load.
Each instruction in \texttt{f'} uses the same execution unit (EU) as the target.
This must be a non-pipelined EU, so that a gadget instruction being issued to the EU blocks a target instruction from issuing.

Figure~\ref{fig:dcu}(b) shows the attack timeline. The value \texttt{z}, which the target address \texttt{A} depends on, takes
$Z$ cycles to compute.
$Z$ is such that before \texttt{z} gets computed, there is enough time for the attack's access load to read the secret,
forward it to the interference gadget, and for the gadget's transmitter load to return (if it is a cache hit).

\noindent \circled{1} The transmitter load accesses a secret-dependent cache line.  This load executes under invisible speculation
protection, but it still retrieves the data from some level of the memory hierarchy.  The attacker can therefore orchestrate
for its execution time to be secret-dependent, by appropriately priming the cache prior to the attack.

\noindent \circled{2} If $\mathtt{secret = 1}$, the transmitter load returns quickly, just before the value \texttt{z} is produced. This makes
the instruction sequence \texttt{f'} in the gadget ready, and its first instruction $f'_1$ is issued to the EU before $f_1$,
the first instruction in \texttt{f}.  Thus, when $f_1$ becomes ready, it is blocked from using the EU.  When $f_1'$ completes,
$f_1$ is issued (due to age-ordered scheduling). However, once $f_1$ completes, $f_2$---which depends on $f_1$---does not
immediately become ready, due to $f_1$'s writeback delay. In contrast, $f'_2$, which depends only on the transmitter,
is already ready and so is issued to the EU.  This creates a cascading effect in which each instruction $f_i$
gets delayed, delaying the target address' computation until the mis-speculation is squashed.

\noindent \circled{3}
If $\mathtt{secret}=0$, the transmitter load does not return before \texttt{z} is produced.
(In delay-based invisible
speculation designs~\cite{sakalis_isca19}, the load is never executed.  In other designs~\cite{invisispec}, the load simply takes a long time to return compared to
the hit case.)  As a result, the target address is computed before the gadget's interfering instructions execute
(if they execute), and the victim load can issue.

\para{$\mathbf{G^I_{RS}}$: Delay \underline{i}nstruction fetch with \underline{RS} contention.} \label{para:icu}
This is a Type 1 gadget that fills the RS for a specific EU, which causes head-of-line blocking in the dispatch queue and forces the Fetch unit to stop I-Cache accesses for fetching instructions.
Figure~\ref{fig:icu}(a) shows the gadget; the target is an instruction fetch by the frontend, so does not appear in the code.
(The gadget's goal is to change when or if the target instruction is fetched.)
The gadget consists of a load (transmitter) and a long sequence of arithmetic (\texttt{ADD}) instructions that depend on the load.
Figure~\ref{fig:icu}(b) shows the attack timeline.

\noindent \circled{1} The transmitter load accesses a secret-dependent cache line, which is setup to hit or miss in the cache hierarchy, depending on the secret.

\noindent \circled{2} If $\mathtt{secret}=1$, the transmitter load is a miss in the D-Cache.  The dependent \texttt{ADD} instructions are fetched and fill up the RS slots, but do not issue. This leads to the RS getting filled up. This consequently creates pressure on the Fetch Unit and the frontend stalls.
Hence, the target instruction is not fetched.  Once the branch resolves, execution continues and the target instruction gets fetched.

\noindent \circled{3} If $\mathtt{secret}=0$, the transmitter load hits in the D-Cache.  Its output is then quickly available to the dependent \texttt{ADD} instructions, which issue as soon as EU resources are available, hence freeing up the RS slots. Since the RS does not fill up, the frontend does not stop fetching. Consequently, the cache line holding
the target instruction is fetched into the I-Cache.

\subsection{From Timing to Cache State Changes} \label{sec:exploit}

We now show how to transform the basic attack primitive (\cref{sec:framework}), which creates
a secret-dependent delay for an unprotected victim memory access, into a cache covert
channel. The insight here is that we can transform a secret-dependent \emph{timing} change---the delay in
the victim's access---into a secret-dependent \emph{cache state} change, by using the delay to
\emph{reorder} the victim access with another (unprotected) \emph{reference} memory access, which occurs at a fixed,
secret-independent time.  Conceptually, the reference access acts as a kind of ``clock,'' helping the attacker to
observe whether the victim load issues before or after some point in time.

Crucially, the only property required from the reference memory access is that its issue time does not depend on the secret.
In particular, the reference access can be issued by the victim or by the attacker, depending on the specific attack.
In what follows, we denote the victim memory access load A (which has address \texttt{A}) and the reference memory access load B (which has address \texttt{B}).
Then, we arrange for load A (A for short) to be issued before or after load B (B for short), depending on the value of the secret.
More precisely: 
The cache state, $\sigma$, is determined by the
sequence of memory accesses to the cache, $\alpha$. 
We assume that $\sigma$ is not commutative, i.e.,
that
$\sigma$($\alpha \, \text{A}\, \text{B}$)$ \neq \sigma$($\alpha \, \text{B} \, \text{A}$).
Formally, therefore, making the order in which \texttt{A} and \texttt{B} access the cache secret-dependent makes the cache
state secret-dependent, creating a cache covert channel.

Non-commutativity of cache-state updates holds for most cache architectures 
as long as both memory accesses target different cache line addresses that map to the same cache set.
For example, the memory access order impacts the set's \emph{replacement state} (e.g., LRU bits), and can be observed by inducing
evictions and monitoring which lines get evicted (by timing memory accesses).

Blocking replacement state-related leakage is explicitly in the scope of invisible speculation (e.g.,~\cite{invisispec,sakalis_isca19,conditional_spec}).  However, we are not aware of such attacks being
demonstrated in practice.
\footnote{Recent work~\cite{LeakingLRU} shows information leakage through cache LRU states, but its channels rely
on more than the ordering of two accesses.}
In \cref{sec:attacks}, we demonstrate a covert channel based on the ordering of two LLC accesses on a commercial
CPU with a sophisticated replacement policy.  Thus, for the following discussion, we assume that achieving
secret-dependent cache access order is equivalent to forming a covert channel.

\subsubsection{Completing the Attacks}
\label{sec:complete}

We now combine the speculative interference gadgets (\cref{sec:gadgets}) with various
types of reference memory accesses to obtain several complete attacks on different points in the invisible
speculation design space.  Each attack creates a cache covert channel by making the secret determine the order
of two unprotected LLC accesses, which may be a victim data access ($V^D$), victim instruction fetch ($V^I$),
or an attacker data access ($A^D$).
(We use $V$ and $A$ to specify whether the victim or attacker thread, respectively, performs the access.)
Table~\ref{tab:complete-attacks} summarizes which defenses are vulnerable to which
attack combinations.

\begin{table}[t]
    \centering
    \small
    \caption{\small \textbf{ Invisible speculation designs vulnerability matrix.}}
    \begin{tabular}{l|p{0.4\columnwidth}p{0.16\columnwidth}p{0.15\columnwidth}}
    \toprule
                            & \multicolumn{3}{c}{\textbf{Accesses With Secret-Dependent Order}} \\
    \textbf{Gadget}         & \multicolumn{1}{c}{$\mathbf{V^D}$-$\mathbf{V^D}$ \& $\mathbf{V^I}$-$\mathbf{V^D}$}   & \multicolumn{1}{c}{$\mathbf{V^D}$-$\mathbf{A^D}$}  & \multicolumn{1}{c}{$\mathbf{V^I}$-$\mathbf{A^D}$} \\
    \midrule
    $\mathbf{G^D_{NPEU}}$   & InvisiSpec (Spectre), DoM (non-TSO), SafeSpec (WFB) & \multicolumn{1}{c}{\textbf{All}}                            & \multicolumn{1}{c}{\textbf{All}} \\
    $\mathbf{G^D_{MSHR}}$   & InvisiSpec (Spectre), SafeSpec (WFB)                & InvisiSpec, SafeSpec, MuonTrap & InvisiSpec, SafeSpec, MuonTrap \\
    $\mathbf{G^I_{RS}}$     & \multicolumn{1}{c}{\textbf{--}}                               & \multicolumn{1}{c}{\textbf{--}}                             & InvisiSpec, DoM \\
    \bottomrule
    \end{tabular}
    \label{tab:complete-attacks}
\end{table}

\para{$V^D$-$V^D$ ordering.}
This attack targets invisible speculation designs that may have multiple unprotected loads executing concurrently.
For example, InvisiSpec and SafeSpec have modes that only defend against control-flow mis-speculation.
In these modes, any load that becomes ready to execute when there are no unresolved branches older than it in the ROB,
performs an unprotected access~\cite{invisispec,safespec}.  A similar case exists with DoM on architectures with a non-TSO memory
consistency model.  In this case, any load can execute without protection if all older branches have resolved and all older stores
and loads have their addresses resolved~\cite{sakalis_isca19}.

We show how to base the attack on the $\mathbf{G^D_{MSHR}}$ or $\mathbf{G^D_{NPEU}}$ gadgets, by modifying the gadget's interference
target so that the victim load A is followed (in program order) by a retirement-bound reference load B,
whose issue time is not affected by the gadget.
Due to space constraints, we fully describe the attack based on the $\mathbf{G^D_{NPEU}}$
gadget; the $\mathbf{G^D_{MSHR}}$-based attack is similar.
Figure~\ref{fig:fulldcu} shows the modified target and the original
gadget (Figure~\ref{fig:dcu}).
Both A and B's address generation depend on \texttt{z}.  If $\mathtt{secret} = 0$ (i.e., no speculative interference), load A accesses
the D-Cache before the reference load B, since the sequence of instructions that generates B,
\texttt{g(z)}, takes longer to
complete than \texttt{f(z)}. However, if $\mathtt{secret}=1$, there is speculative interference, so A's generation is delayed
while B's is not, and load B accesses the D-Cache first.

\para{$V^I$-$V^D$ ordering.}
Modifying the target in the $V^D$-$V^D$ attack so that the branch condition \texttt{N} depends on load A makes the delay of load A also delay the branch's resolution time, i.e., when the squash occurs.  This can change the order of a post-squash instruction fetch---which is unprotected, as it is of the correct execution path---with respect to load B.

\begin{figure}[t]
\begin{lstlisting}[basicstyle=\footnotesize,xleftmargin=0.3cm]
z = ...   // takes Z cycles
A = f(z)  // takes F cycles
y = load(A)
B = g(z)  // takes G > F cycles
v = load(B)
if (i < N): // mispredict taken (miss on N)
   secret = load(&TargetArray[i])
   // Interference Gadget
   x = `load(&S[secret * 64])` // secret=1->hit, secret=0->miss
   f$'$(x)
\end{lstlisting}
\caption{\small \textbf{ Reordering victim loads by exploiting contention on a non-pipelined EU.
Instruction sequences \texttt{f} and \texttt{f'} use the same non-pipelined EU. Instruction sequence $g$ uses a different EU.}}
\label{fig:fulldcu}
\end{figure}

\para{$V^D$-$A^D$ ordering.}
Many invisible speculation designs unprotect a load only when it becomes the oldest load or the oldest instruction
in the ROB.  This is the case in
InvisiSpec's Futuristic mode~\cite{invisispec}, SafeSpec's wait-for-commit mode~\cite{safespec}, Conditional
Speculation~\cite{conditional_spec}, and MuonTrap~\cite{muontrap}.  These designs make it impossible to reorder
unprotected victim loads, as no two such loads can execute concurrently.  As noted above, however, the
same effect---secret-dependent order---can be achieved if the attacker performs the reference access.
For this, the attacker simply needs to issue an LLC access to the same set accessed by the $V^D$ load from another core,
at a fixed time after inducing the mis-speculation. This attack can be based on either of the $\mathbf{G^D_{MSHR}}$ or $\mathbf{G^D_{NPEU}}$ gadgets.

\para{$V^I$-$A^D$ ordering.}
As in the $V^I$-$V^D$ case, the $\mathbf{G^D_{MSHR}}$ and $\mathbf{G^D_{NPEU}}$ gadgets can be used to target the branch condition,
delaying a post-squash instruction fetch on the correct execution path. This can be measured using the attacker's LLC access as a reference clock. In contrast, the $\mathbf{G^I_{RS}}$ gadget only impacts the timing of instruction fetches in the mis-speculated path. Hence, the delay it introduces
for instruction fetches can only be observed if I-Cache accesses are not protected by the invisible speculation scheme, as in
InvisiSpec and DoM.

\para{Attack landscape summary}
\emph{Every} invisible speculation design
we have evaluated
is vulnerable to at least one of the attacks described above.
Table~\mbox{\ref{tab:complete-attacks}} summarizes which designs are vulnerable to which attack combinations.
The differences in security manifest in whether an attacker can reorder unprotected victim accesses or must rely on its own access as a ``reference clock.''

\subsection{Existence of Interference Gadgets/Targets (Senders)}

We refer to the combination of an interference gadget and a target as a \emph{sender}, i.e., the sending side of the cache covert channel.
It is natural to ask if senders exist in ``the wild,'' given their specific structure.
There are several real-world attack settings~\cite{intel-taxonomy} in which the attacker has some control over the instruction stream and can craft senders.  These settings include (1) the \emph{in-domain} setting, where a software sandbox executes attacker-controller code, as in the case of in-browser JavaScript code or user-supplied Linux eBPF kernel extensions~\cite{eBPF}; and (2) the \emph{domain-bypass} setting, where the attacker runs its own program, attempting to use its mis-speculated execution to steal secrets from another hardware protection domain, e.g., Meltdown~\cite{meltdown}.
Finding whether speculative interference senders exist in victim programs is interesting future work.

Conceptually, however, even the fact that senders \emph{might} exist creates uncertainty about security
on an invisible speculation system. Users and developers cannot \emph{know} if their program contains a sender without performing
program analysis to verify their non-existence. Having to rely on such analysis to guarantee security undermines the efficacy of invisible speculation as a software-transparent hardware defense.

\section{Attack Demonstrations} \label{sec:attacks}

In this section, we demonstrate concrete proof-of-concept (PoC) speculative interference attacks based on the ideas from \cref{sec:break} on a commercial machine.
Although invisible speculation schemes are not implemented today, we can emulate their behavior by arranging for loads that would be made `invisible' to return data in secret-dependent amounts of time.
At the same time, by evaluating on real hardware, we must address many details in real machines that are simplified in simulators (e.g., LLC replacement policies, RS limits).

We evaluate multiple D-Cache PoCs and a variation of the I-Cache PoC described in \cref{sec:gadgets} namely, $\mathbf{G^D_{NPEU}}$, $\mathbf{G^D_{MSHR}}$ and $\mathbf{G^I_{RS}}$.
All the PoCs were successfully implemented and the attacks successfully leak secret bits to the attacker.
We only show the $\mathbf{G^D_{NPEU}}$ and $\mathbf{G^I_{RS}}$ attacks for space and refer to them as the D-Cache PoC (\cref{sec:dcu:poc}) and I-Cache PoC (\cref{sec:icu:poc}) respectively.
Of independent interest, our D-Cache PoC requires constructing a novel receiver able to read changes in replacement state for the QLRU\_H11\_M1\_R0\_U0 replacement policy (\cref{sec:dcu:receiver}).
All attacks change cache state, with a receiver (attacker) that monitors execution from another physical core (CrossCore; ~\cref{sec:threatmodel}).

\subsection{Methodology} \label{sec:meth}

\textbf{Processor details.}
We evaluate on an Intel Core i7-7700 Kaby Lake CPU with 4 physical cores 
running at a base frequency of 3.6GHz, with hyper-threading enabled.
Each core has a unified reservation station, that is shared across execution units, stores up to 97 micro-ops, and has 
8 execution unit ports (numbered 0 through 7).
Each core has two levels of private cache (a 32KB L1-instruction and 32KB L1-data cache, 256KB of combined L2)
and 8MB of Shared L3 (LLC) cache~\cite{inteldocs,wikichip:kaby}.

\noindent
\textbf{Tools borrowed from prior work.}
We trigger branch mispredictions by training the target branch in a given direction (similar to \cite{spectre}).
Likewise, we delay branch resolution by having the branch predicate be the result of a pointer chase.
The attacks also use a Flush+Reload-style~\cite{flush+} receiver.

Finally, the D-Cache PoC uses standard techniques to construct eviction sets in the LLC~\cite{last_level_cache_practical}, which are 
sets of cache lines that map to the same LLC set in the same LLC slice.
By accessing lines in an eviction set, the attacker can efficiently evict other lines whose set and slice is known.

\subsection{D-Cache PoC}
\label{sec:dcu:poc}

Recall from \cref{sec:gadgets}, the key principle in the $\mathbf{G^D_{NPEU}}$ attack is for the attacker to observe the reordering of two bound-to-retire loads.
Our PoC measures this ordering by mapping the two loads to the same LLC set and measuring changes in replacement state.

To deploy the attack there are two ingredients that need to be developed.
First (\cref{sec:dcu:loadorder}), an implementation of the $\mathbf{G^D_{NPEU}}$ sender, i.e., to reorder older bound-to-retire loads.
Second (\mbox{\cref{sec:dcu:receiver}}), a novel receiver capable of measuring differences in LLC replacement state.
We consider both of these to be of independent interest, i.e., to reorder older non-load instructions to perform different speculative interference attack variants or to be used in entirely different attack settings (in the case of the replacement state-based receiver).

\subsubsection{Sender (Load Reordering)}
\label{sec:dcu:loadorder}

We use the same notation to denote the victim load A and reference load B as in~\cref{sec:exploit}.
To reorder the loads to two addresses \texttt{A} and \texttt{B}, we follow the structure from Figure~\ref{fig:fulldcu}. Namely, there are two sequences of instructions, \texttt{f(z)} and \texttt{g(z)}, that generate addresses \texttt{A} and \texttt{B} respectively.
An interference gadget only affects \texttt{f(z)}. In presence of the gadget, load A is delayed to issue after load B whereas regularly it would issue before load B.

First, we consider the address generation for load A and the interference gadget in isolation.
We implement \texttt{f(z)} and \texttt{f'(x)} (Figure~\ref{fig:fulldcu}) as repeated sequences of same instructions, called the \emph{target instruction} and \emph{gadget instruction}, respectively.

We pick suitable instructions (i.e., that maximize the interference of the gadget on the target) as follows.
We identify high latency, low-throughput instructions that use the same execution port.
Low-throughput allows for an issued instruction in the interference gadget
to block the execution port of ready-to-schedule
instructions in the interference target;
high latency maximizes the time it blocks instructions in the interference target. 
Finally, the gadget instruction should be composed of only a few micro-ops.
This allows more
instructions in the interference gadget 
to occupy RS simultaneously,
which increases the likelihood of them getting issued concurrent to the target instructions.

\vspace{-1.5ex}
\begin{figure}[hbt!]
    \centering
    \includegraphics[width=0.95\linewidth]{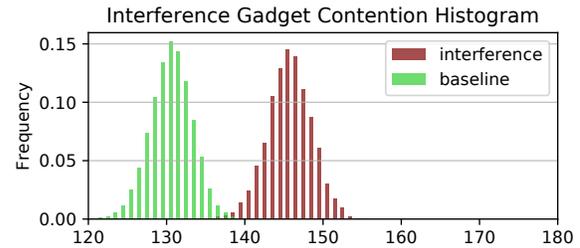}
\caption{\small \textbf{
The average time (measured with a clock thread~}\cite{schwarz2017fantastic, schwarz2017malware}) to execute the interference target changes by $\sim16$ clock ticks
(80 rdtsc cycles)
based on the presence or absence of the interference gadget.}
\label{fig:dcuload}
\end{figure}
\vspace{5pt}

Based on the above process, our PoC uses the \texttt{VSQRTPD} instruction for both gadget and target.
\texttt{VSQRTPD} consists of only 1 micro-op executed on the
core's execution port 0
and has observed latencies of 15--16 cycles and reciprocal throughput of 9--12 cycles~\cite{fog2011instruction}.
We also verified that the attack is functional with \mbox{\texttt{VDIVPD}}.
Figure~\ref{fig:dcuload} shows the time from the issue of the first instruction  of \texttt{f(z)} to the completion of the load A in the presence (interference) and absence (baseline) of the interference gadget's execution. 
The takeaway is there is a clear timing difference in the
interference
target's
execution depending on presence/absence of the
interference
gadget's
execution. This is the secret-dependent delay imposed by the gadget on the victim load.

\subsubsection{Receiver (Monitoring Replacement State)}
\label{sec:dcu:receiver}

With the capability to reorder two loads, the next ingredient for the attack is to translate a reordering of loads into a persistent cache-state change.
We
achieve
this using the cache replacement state.\footnote{RELOAD+REFRESH~\cite{briongos2020reload+} also uses replacement-state manipulation principles to execute a cache-based attack. The distinction in this work is we try to identify the victim's load issue order, whereas they try to identify the presence of a victim's access to a target address.}
For the rest of the section, we use the notation A-B to indicate the order in time in which the loads are issued,
i.e., A-B means A issues first and  vice-versa.
We also assume access to eviction sets (EV; \cref{sec:meth}). 

Our attack targets the replacement state because we are only changing the order of loads.
Changing the order of loads is different than changing which loads are issued as in a normal cache-based attack.
For example, a standard LLC Prime+Probe attack, without a very fine probe granularity, would observe both A and B in the cache, regardless of their order and be unable to distinguish A-B from the B-A case.

Translating load issue order into a persistent replacement state change is not difficult in textbook replacement policies, such as LRU, as the ordering directly influences replacement priority ranking.
However, replacement policies in modern machines, such as our target processor, are more complex.
The new technical challenge for the attacker is that fresh insertions of A and B are ranked equally.

This new challenge can be overcome by providing a technique to extract replacement state data from the replacement policy on the Kaby Lake machine.
To identify the replacement policy on our machine, we used a CacheAnalyzer tool by nanoBench~\cite{abel2019nanobench}. The resulting replacement policy is approximately QLRU\_H11\_M1\_R0\_U0 (``Quad-age LRU'') on specific cache sets~\cite{vila2019cachequery}.\footnote{It is likely the case that the LLC cache sets do not strictly abide by this replacement policy and have an adaptive replacement policy. However, for the purposes of this PoC, the attack strategy that creates observable replacement state changes on QLRU\_H11\_M1\_R0\_U0, also creates observable replacement state changes on our machine.}
QLRU is a Static-RRIP Replacement policy variant with a 2 bit field used for the age of a cache line~\cite{jaleel2010high,abel2019nanobench}, summed up here:

\begin{itemize}[nosep,leftmargin=1em,labelwidth=*,align=left]
\item M1: Insertion policy. Inserts cache lines with age 1.
\item H11: Hit promotion policy. Promotes a line of age 3 to age 1, age 2 to age 1, and age 1/0 to age 0 upon hit.
\item R0: Eviction policy. Insert to leftmost location if cache set is not full; otherwise, evict block corresponding to the leftmost physical tag with age 3.
\item U0: Age update policy. Increments age fields of all cache lines until there is a candidate ready for eviction (age = 3).
\end{itemize}

\para{Attacker Receiver Protocol.}
We now describe how the attacker decodes from the replacement state whether A-B or B-A occured.
At a high level, similar to a traditional cache attack, the attacker thread first primes the LLC set, waits for the victim to issue its secret-dependent ordering, and finally probes the LLC set to determine which ordering the victim issues.
Due to the nature of QLRU, however, the details are different from conventional attacks.
Specifically, the attacker first constructs two eviction sets of size 
LLC\_ASSOCIATIVITY-1
elements, call these EVS1 and EVS2,  which map to the same LLC set and slice as A and B. The attacker then uses the following access sequences to prime and probe the cache set:

\begin{itemize}[nosep,leftmargin=1em,labelwidth=*,align=left]
    \item \emph{Prime Sequence: Access EVS1 many times + Access A}
    \item \emph{Probe Sequence: Access EVS2}
\end{itemize}

The attacker accesses 
EVS1 many times in order to saturate their age at 0, leaving A with an age of 3.
To be able to access address \texttt{A}, our current PoC requires that the receiver share memory with the victim (hence the use of Flush+Reload).

\begin{figure}[hbt!]
    \centering
    \includegraphics[width=0.95\linewidth]{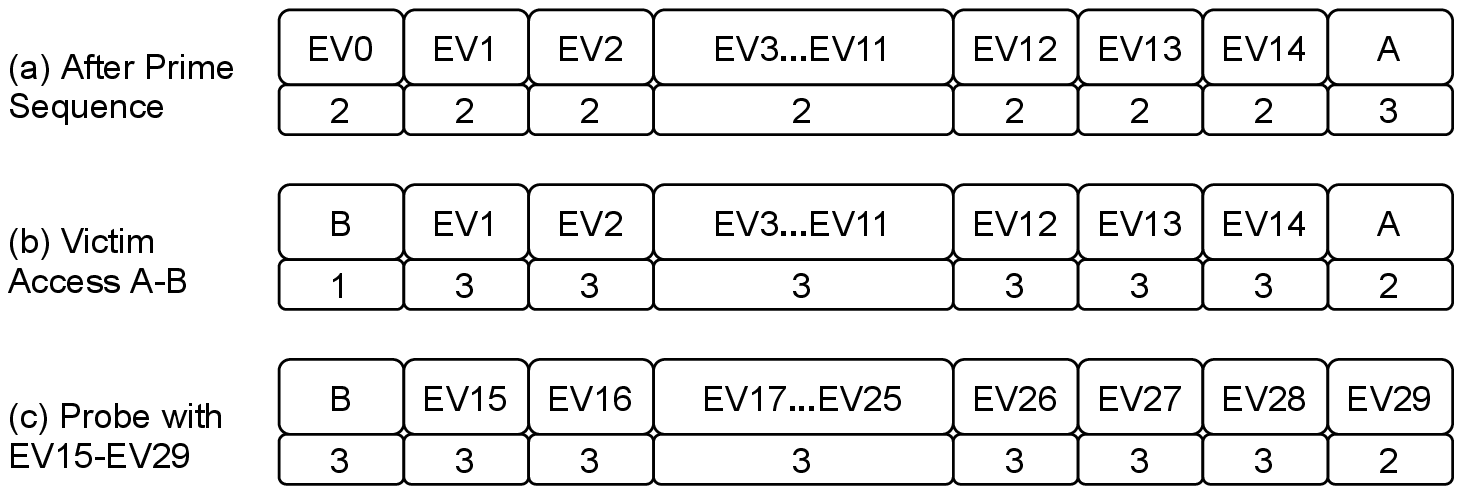}
\caption{\small \textbf{QLRU State for the targeted cache set. EV\textit{N}, A, B represent addresses and numbers represent the age for each cache line. (a) shows the cache state after attacker primes the cache. (b) \& (c) represent the cache states after the victim runs (with pattern A-B) and after the attacker completes the probe. A victim access pattern of B-A has analogous state changes.}
}
\label{fig:qlru}
\end{figure}

For our machine, the targeted cache sets are 16-way associative.
We will refer to elements in EVS1 as EV0-EV14, and elements in EVS2 as EV15-EV29.
The resulting cache states for prime and probe with the A-B sequence is displayed in Figure~\ref{fig:qlru}. The main idea is that only A or B is still resident in the LLC by the end of
prime+victim\_accesses+probe sequence.

\subsubsection{End-to-End Attack}
\label{sec:dcu:endtoend}

\begin{figure}[hbt!]
    \centering
\includegraphics[width=\columnwidth]{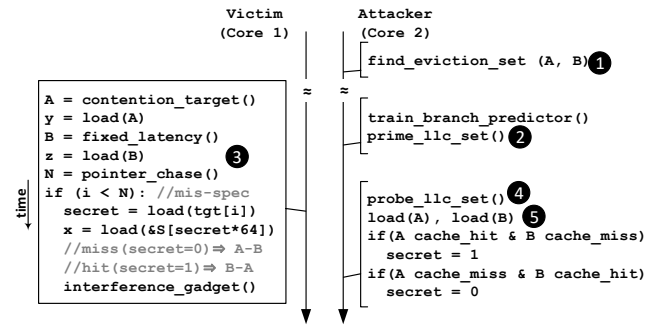}
\caption{\small
\textbf{An End to End visualization of the D-Cache attack.}
}
\label{fig:dcu:end_to_end}
\end{figure}

In this section, we present the overall D-Cache PoC. 
The attack steps shown in Figure~\ref{fig:dcu:end_to_end} are explained in detail below:

\noindent \circled{1}
Attacker initializes eviction sets
based on addresses \texttt{A}, \texttt{B}.

\noindent \circled{2}
Attacker primes the LLC set replacement state (\cref{sec:dcu:receiver}) and mis-trains the victim's branch predictor.

\noindent \circled{3} Victim issues loads A and B, where order depends on the secret (\cref{sec:gadgets}).
If secret = 0, A-B is issued, and if secret = 1, B-A is issued.

\noindent \circled{4}
Attacker probes the LLC set replacement state (\cref{sec:dcu:receiver}) and observes the residency of lines for addresses \texttt{A} or \texttt{B} in the LLC set. The residency is determined by issuing a timed access to address \texttt{A} and \texttt{B} and comparing it with a LLC cache miss threshold.

\noindent \circled{5}
Attacker attempts to identify the secret bit. If the victim issues the load sequence A-B (secret = 0), the expectation is for load A to be a cache miss and load B to be a cache hit. If the victim issues the load sequence B-A (secret = 1), the expectation is for load A to be a cache miss and load B to be a cache hit. Cases where both accesses are cache misses can happen due to noise and are ignored.

\noindent \circled{6}
Attacker repeats steps 2-5 as needed to increase confidence.

Note, while our PoC observes load-reordering through replacement state, other receivers not based on replacement state might be possible.

\subsection{I-Cache PoC} \label{sec:icu:poc}

We now describe a PoC for the $\mathbf{G^I_{RS}}$ attack from \cref{sec:gadgets}. 
The attack works by creating contention on available reservation stations to create a ripple effect that eventually stops the frontend from fetching more instructions, causing changes to the I-Cache access pattern.

\subsubsection{Experiment Setup} \label{sec:icu:setup}

As with the D-Cache PoC, we show the I-Cache PoC given an attacker that monitors state from another physical core through the LLC. 
The pseudocode for the victim is described in Figure~\mbox{\ref{fig:icu}}. 
Without loss of generality, the target instruction used at \mbox{\cref{line:tgti}} in Figure~\mbox{\ref{fig:icu}} is a shared library function call. 
For simplicity, our attack slightly differs from that explained in Figure~\mbox{\ref{fig:icu}}.
Specifically, we move the target instruction into the mis-speculated path (before the branch join point).
Thus, in a correct execution, the target instruction will not be fetched (as opposed to fetched later).
The receiver (attacker) on the adjacent physical core issues a load to a shared library function 
to perform the reload step.

\begin{figure}[hbt!]
    \centering
    \includegraphics[trim={0 0.4cm 0 0.2cm},width=0.95\linewidth,clip]{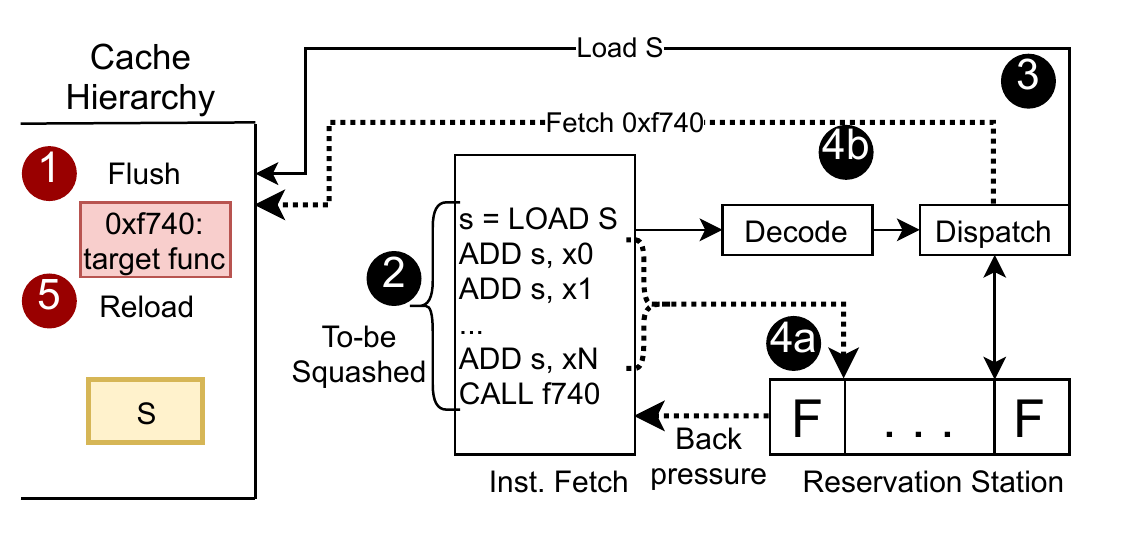}
    \caption{\small 
    \textbf{Steps involved in performing an I-Cache attack. Dotted lines are conditional events depending on Block S's presence in cache. Red 
    represents attacker actions and black 
    represents victim actions.}}
    \label{fig:icu_arch}
\end{figure}
\subsubsection{Attack Details} \label{sec:icu:detail}

Here we refer to the pseudocode in Figure~\mbox{\ref{fig:icu}} and describe the attack sequence. Item numbers refer to steps in Figure~\mbox{\ref{fig:icu_arch}}.

\noindent \circled{1} Attacker first primes the cache hierarchy by flushing out the target address (shared function pointer) from the I-Cache.

\noindent \circled{2} When the victim runs next, it mis-speculates on a branch (\cref{line:cond} in Figure~\ref{fig:icu})
that prompts the frontend to fetch transient (bound-to-squash) instructions, which
dispatches a secret-dependent load instruction at address \texttt{S}, followed by a large number of 
arithmetic instructions dependent on the loaded value \texttt{s}.

\noindent \circled{3}  The next steps proceed similarly to what we describe in the \cref{sec:gadgets} $\mathbf{G^I_{RS}}$ paragraph. The victim dispatches the secret-dependent load (Load S in Figure~\ref{fig:icu_arch}, \cref{line:load} in Figure~\ref{fig:icu}), which is setup to hit or miss in the cache hierarchy based on a secret value \texttt{S}.

\noindent \circled{4a} Miss: Load S misses, creating a frontend stall due to dependent ADD instructions. Hence, the target instruction (Call \texttt{0xf740} in Figure~\ref{fig:icu_arch}) is not fetched.  When the branch resolves, execution continues but because the target instruction was on the mis-speculated path, the target address was never fetched into the cache.

\noindent \circled{4b} Hit: Load S returns quickly and no RS congestion occurs. The target instruction is executed and hence the target address cache line is fetched into the I-Cache. Since the branch resolves after the target instruction is executed, the fetched line leaves a persistent change in cache state after the mis-speculated instructions are squashed.

\noindent \circled{5} After waiting for the victim to run, the attacker performs a standard probe step, re-loading the shared function pointer, either via a function call (I-Cache) or by loading the contents of the function pointer (D-Cache).

\subsection{Attack Evaluation} \label{sec:ev_res}

\begin{figure}[hbt!]
    \centering
    \includegraphics[width=1.0\linewidth]{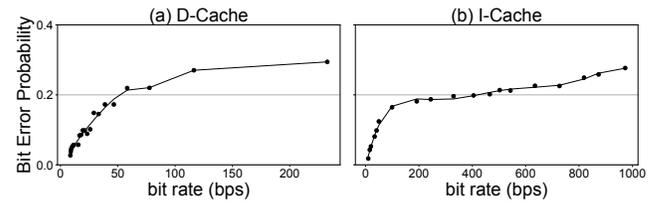}
\caption{\small
{\bf
Attack PoC channel error vs. bit rate.
(a) DCU PoC (\cref{sec:dcu:poc}),
(b) ICU PoC} (\cref{sec:icu:poc}).
}

\label{fig:poc_results}
\end{figure}
\vspace{.05in}

We run the PoCs in a cross-core setting and evaluate the end-to-end covert channel error rate vs. throughput in 
Figure~\mbox{\ref{fig:poc_results}} (a) and (b) for the D-Cache and I-Cache PoCs, respectively.
Throughput is defined as the number of secret bits transmitted per unit time.
It is represented as bits per second (bps) and evaluated by measuring the CPU cycles required to leak 1 bit.
Error rate is defined as the number of incorrectly inferred bits over the total number of bits transmitted.
We can trade-off error rate and bit rate by changing PoC parameters, e.g., the number of times the PoC is run to leak each bit, or the amount of time spent trying to mistrain the branch predictor.
We note the I-Cache PoC has a higher transmission rate than the D-Cache PoC because reordering loads in the D-Cache PoC has a higher chance to fail due to noise in the system.

\section{Defenses}
\label{sec:defense}

We now discuss various approaches for invisible speculation designs to block speculative interference attacks.
To this end, we first propose a formal definition of what it means to block all cache covert channels (\cref{sec:goal}).
We describe two designs that achieve this goal.  The first design  (\cref{subsec:basic}) 
is straightforward, but imposes significant performance overhead, unlike current designs ~\cite{sakalis_isca19,muontrap}. We thus propose a high-level
approach for a more efficient solution (\cref{advanced_defense}), whose exploration we leave to future work.

\subsection{Ideal Invisible Speculation} \label{sec:goal}

We define an \emph{ideal invisible speculation} security property, which formally models the security goal
of eliminating all speculative execution-induced cache covert channels.  Informally, ideal invisible speculation requires
that the system's cache state is invariant of speculative execution.

More formally: We assume a multi-core system with private L1 I- and D-Caches and a shared L2 cache (or L2).
In an invisible speculation design, the L2 can receive visible and invisible accesses.  A \emph{visible}
access corresponds to a standard cache fill or writeback, causing changes in both the L1 and L2.
An \emph{invisible} request is a request type added by the defense; it does not cause state changes in the L2 and its
response does not change state in the L1.  We assume that the attacker sees the sequence (without timing information)
of visible L2 accesses.  We call this the \emph{L2 access pattern.}

We formulate the security goal of the L2 access pattern being invariant of speculation as follows.
Given an execution $E$ of the microarchitecture, define $C$($E$) as the L2 access pattern in $E$.
Define $NoSpec$($E$) as the execution that would have occurred if $E$ had no mis-speculations.
Then \emph{ideal invisible speculation} is the following property, akin to non-interference~\cite{noninterference}:
For any execution $E$: $C$($E$)$=C$($NoSpec$($E$)).

\subsection{Basic Defense Design}
\label{subsec:basic}

Here, we present a simple solution that can provide ideal invisible speculation.
The idea is that, when instructions that might cause a mis-speculation are inserted
in the ROB, the hardware automatically inserts a special type of fence. The fence
allows subsequent instructions to be inserted into the ROB but prevents them from being issued until the instruction before the fence becomes non-speculative.

To achieve ideal invisible speculation, fences must be inserted after any instruction that
may cause a squash. This threat model (considering all forms of speculation) is sometimes
referred to as the \emph{Futuristic} model~\cite{invisispec}. The design can be tuned
to consider only control-flow speculation (the \emph{Spectre} model~\cite{invisispec})
by placing fences only after branches. This requires adjusting the security property
to consider only control-flow speculation.

\para{Evaluation of the Basic Defense.} We evaluate the performance of the basic defense design on
the Gem5~\cite{gem5} simulator.
We model a high performance multi-core system (8-issue OoO 2\,GHz cores
with 32\,KB L1 D-Cache and 2\,MB per-core shared L2 banks).
This configuration is similar to systems used in previous invisible
speculation work (e.g.,~\cite{invisispec,sakalis_isca19,muontrap}).
We use the SPEC CPU2017~\cite{spec} benchmarks with reference input size
and Simpoints~\cite{simpoints} to identify around 10 representative execution
regions per benchmark and run 10 million instructions per simpoint.

\begin{figure}[t]
    \centering
    \includegraphics[width=1\columnwidth]{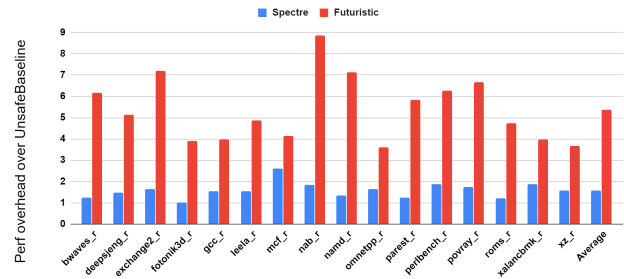}
    \caption{\small \textbf{ Performance of basic defense on SPEC2017 benchmarks.}
    }
    \label{fig:result}
\end{figure}

Figure~\ref{fig:result} shows the performance overhead of the basic defense
design over the unsafe, unmodified processor, under both Spectre and
Futuristic threat models.
When adding the basic defense scheme, the execution time becomes on average 1.58$\times$ the execution time of the unsafe baseline for Spectre threat model.
In terms of the Futuristic threat model, the execution time of the
basic defense scheme is on average 5.38$\times$
the execution time of the unsafe baseline.
In comparison, prior works report overheads of $<5$\% and $<20$\% in the
Spectre and Futuristic models, respectively~\cite{invisispec,sakalis_isca19,muontrap}.
Therefore, while the simple solution can achieve ideal invisible speculation,
it does so at a dramatic performance cost.

\subsection{Discussion: Potential Advanced Defense}
\label{advanced_defense}

A high level principle for achieving ideal invisible speculation is: \emph{a speculative instruction must not influence the execution of a non-speculative instruction}.
This principle does not preclude an instruction from speeding up younger instructions,
which is the basis for invisible speculation's performance benefits. It does imply the
microarchitectural rule of blocking loads from changing cache state while speculative, as that can affect non-speculative
instructions. As we have shown, however, more microarchitectural rules are required to realize this principle,
creating a design space to explore.

One possible approach is based on the following two rules:
(1) no instruction ever influences the execution time of an older instruction,
and
(2) any resources allocated to an instruction at the interface of the frontend and the execution engine are not deallocated until the instruction becomes non-speculative.
Rule (1) is straightforward to obtain if every microarchitectural resource is perfectly pipelined but requires further research to implement otherwise.
Rule (2) makes instruction fetch rate invariant of speculation, which guarantees that speculation cannot influence when instructions
that eventually retire begin executing.

\para{Not Influencing Older Instructions.}
This rule can be implemented by assigning a priority tag to each instruction, based on its age in the ROB.
The general strategy follows three steps. First,
when two instructions with priorities $i$ and $j$ are about to use
any shared resource, the hardware gives precedence to  the instruction with higher priority. Second, to prevent counter wrap-around problems,
we maintain two priority tags (head and tail) and use logic akin to FIFO full/empty to check priority tag values on instructions.
Third, when a branch is squashed, the priority tag is reset to the correct~value.

This design is straightforward when every single resource is perfectly pipelined. This is currently the case for pipelined EUs, cache ports and banks, and writeback links. However,
for resources that are not perfectly pipelined (e.g., non-pipelined EUs), three different choices are available.

One approach is to
make them fully pipelined (at the expense of longer latency).
A second approach is to make the corresponding resource scheduler smarter, so that it ``looks ahead in time'' and anticipates if, by assigning the resource to a low-priority instruction now, one may have to stall a higher-priority instruction later. In this case, the low priority instruction is stalled until the higher priority one uses the resource. This strategy may not be possible all the time (or be quite expensive/slow).
The third approach is to design the EU to be ``squashable''. This means that it can be freed-up on demand if a higher-priority instruction requests the EU. This complicates the design, as it requires that the instruction currently using the resource be ``re-issuable'' (i.e., the hardware holds its state and can reuse it to relaunch the operation).

\para{Not Deallocating Resources Early.}
This rule requires that a speculative instruction releases the hardware resources
it uses only when it becomes non-speculative or gets squashed. Examples of such resources
are reservation stations and execution units.  This rule makes the duration of time that
the instruction occupies any resource independent of the instruction's operands.
The trade-off is that the instruction may hold on to the resource for longer than in
the baseline design.

\para{Takeaway.}
Extending the invisible speculation approach to block speculative interference attacks appears to involve significant
complexity and efficiency costs.
Whether defenses focusing on only cache attacks---but \emph{fully} blocking them--- can
be simpler or more efficient than defenses with more comprehensive
threat models~\cite{stt,sdo,nda,SpecShield} is an interesting question for future work.

\section{Related Work}

Most speculative execution attacks that have been presented to date build on cache-based covert channels to leak data, being inspired by either Spectre~\cite{spectre, sgx_spectre, spectre_returns, ret2spec, spec_variant_four, wampler2019exspectre} or Meltdown~\cite{meltdown, canella2019fallout, Schwarz2019ZombieLoad, ridl, foreshadow, foreshadow_ng}.
To our knowledge, only SMoTherSpectre~\cite{smother} and NetSpectre~\cite{netspectre} make use of alternative covert channels, such as port contention, for speculative execution attacks.

We provide background on \emph{invisible speculation} schemes~\ISPEC in~\cref{sec:background}.
 CleanupSpec~\cite{cleanup_spec} targets the invisible
speculation goal of blocking only cache covert channels but uses a unique approach of (1) undoing cache occupancy changes upon a squash
and (2) using randomized replacement to block replacement-related leakage.
CleanupSpec does not block speculative interference
but makes its exploitation more challenging.
CleanupSpec relies on DoM or InvisiSpec to protect the I-Cache. Thus, a speculative interference attack that delays branch resolution based on a secret can change the order of victim I-Cache accesses and attacker D-Cache accesses and yield an attack (as in $V^I$-$A^D$ ordering, \cref{sec:complete}). For D-Cache protection, CleanupSpec uses randomized/encrypted address caches, which do not hide whether a previously un-cached line was accessed or not. Thus, a similar attack as in the I-Cache case can go through by delaying a branch resolution, whereby the attacker can change how many D-Cache accesses are made on the mis-speculated path ($V^I$-$V^D$ ordering, \cref{sec:complete}). We leave this as future work.

Beyond invisible speculation, there are several other hardware mechanisms designed to block speculative execution attacks~\cite{stt,oisa,spectre_guard,ConTExT,sdo,nda,SpecShield,dawg,2018ContextSensitiveF}.
Data-oblivious ISA extensions~\cite{oisa}, SpectreGuard~\cite{spectre_guard}, ConTExT~\cite{ConTExT}, DAWG~\cite{dawg} and CSF~\cite{2018ContextSensitiveF} require some degree of software support (e.g., setting up cache partitions, users annotating what data is secret) which severely constrains adoption.
STT~\cite{stt,sdo}, NDA~\cite{nda} and SpecShield~\cite{SpecShield} are software-transparent hardware mechanisms that propose selective speculation, allowing certain instructions to execute speculatively while delaying (or executing in a data-oblivious fashion~\cite{sdo}) others.  
These schemes, in principle, can block speculative interference attacks by delaying speculative instructions beyond loads.
Yet, none can comprehensively defeat all speculative interference attacks.
For example, while STT soundly blocks speculative interference attacks that leak transiently accessed data, it offers no protection against attacks that leak non-transiently accessed (bound-to-retire) data.

Concurrent to this work, Fustos et al.~\cite{spectre_rewind} also observed that younger speculative instructions can influence the timing of older bound-to-retire instructions.
Yet, their SpectreRewind attack is a traditional contention attack (\cref{sec:introduction}) and explicitly outside of the scope of invisible speculation schemes.

\section{Conclusion}
This paper presented speculative interference attacks, which show that invisible speculation schemes are not immune to cache attacks.
The broader implication of our work is to demonstrate the security pitfalls of a well-studied approach to building secure processors, namely to ignore ``bandwidth'' or ``contention'' or ``intermittent'' covert channels and solely focus on cache-based channels.
Specifically, we show how an attacker can \emph{convert} timing changes into persistent-state changes.
Long term, we hope our work helps set a research agenda towards more comprehensive security definitions and more secure, efficient invisible speculation mechanisms.

\begin{acks}
We thank the anonymous reviewers and our shepherd Dmitry Ponomarev for their valuable feedback.
This work was funded in part by NSF under grants 1942888 and 1954521, ISF under grant 2005/17, Blavatnik ICRC at TAU, and by an Intel Strategic Research Alliance (ISRA) grant.
\end{acks}


\appendix
\section{Artifact Appendix}

\subsection{Abstract}

The artifact contains two attack PoCs, the (\texttt{dcache\_poc}) and (\texttt{icache\_poc}) as described in \cref{sec:attacks}. The artifact evaluates the entire attack idea described in the paper in a systematic manner. It contains makefiles, run scripts and detailed README to as a walk-through for each step and to reproduce the result figures described in the main paper. The artifact has been validated on Intel CPUs (i7-7700 and i7-9700) running Ubuntu OS (18.04). The PoC's require a relatively controlled multi-core environment, with two cores dedicated to running attacker and victim snippets; a several-core machine is preferred. Also, as a step in verifying the replacement state policy, there is a requirement of installing kernel modules related to the nanoBench~\cite{abel2019nanobench} workflow.

\subsection{Artifact Check-List (Meta-information)}

{\small
\begin{itemize}
  \item {\bf Algorithm:} Using existing methods to create eviction sets, but implementing specific PoCs developed in the paper
  \item {\bf Program:} D-Cache and I-Cache PoCs
  \item {\bf Compilation:} gcc
  \item {\bf Run-time environment:} Ubuntu 18.04
  \item {\bf Hardware:} Intel i7-7700
  \item {\bf Run-time state:} Sensitive to noisy run-time state, ideally requires low traffic/evictions from the LLC
  \item {\bf Execution:} Pre-profiling for replacement state policy
  \item {\bf Output:} Figures 7 and 11
  \item {\bf Experiments:} Timing difference distributions and leak accuracy rates
  \item {\bf Publicly available:}  Yes 
  \item {\bf Code licenses:}  Apache-2.0 License 
  \item {\bf Data licenses:} None
\end{itemize}
}

\subsection{Description}
\newcommand{\size}[2]{{\fontsize{#1}{0}\selectfont#2}}

\subsubsection{How to Access} 
The PoC's can be accessed from Github:
\url{https://github.com/FPSG-UIUC/speculative_interference}

\subsubsection{Hardware Dependencies}
The proposed attack PoC relies on a proper contention induced by the gadget instructions (Figure~\ref{fig:dcuload}) and the specific replacement algorithm in the LLC (QLRU\_H11\_M1\_R0\_U0). We have tested it on Intel i7-7700 CPU. We expect any processor having the above mentioned replacement state should run the experiment successfully.

\subsubsection{Software Dependencies}
We have tested our PoCs on Ubuntu 18.04 and Ubuntu 20.04. We have seen issues with some of the gcc default compiler options (position independent code) while using Ubuntu 16.04 with the I-Cache PoC. We do not have any specific kernel or package dependencies. For our runs, we used gcc 9.3 with make 4.2.1 and python 2.7.12 for the included scripts.

\subsection{Installation}
No specific installations required, but we do require the use of a run script to set the environment which is provided as \texttt{run-first.sh}.

\subsection{Experiment Workflow}
\paragraph{D-Cache PoC}
Initially we confirm the robustness of the interference gadget used in the PoC. The desired result is a clean difference in the timing distributions of the interference target in presence/absence of the gadget.
Then testing is done to verify the correct replacement state policy (QLRU\_H11\_M1\_R0\_U0) for which the PoC is developed. This was achieved using the nanoBench test suite to converge on the given replacement policy. If an alternative replacement policy is detected, the prime and probe sequences may need to be adjusted.
Once the replacement policy is confirmed and before the PoC can be run, a contiguous buffer in memory is allocated using  shm-crea.c to be used for eviction-set development. The page ids output from shm-crea are stored in a file named "pageids.txt" and are read from the dcache\_poc. 
The PoC itself runs in a series of steps, beginning with eviction set and flush set establishment. Once these have been initialized, a succession of experiments are ran leading up to the multi-core PoC. Further details of running the tests are specified in the README.

\paragraph{I-Cache PoC}
The I-Cache PoC has two variants where the attack PoC can be run as a single core or cross core multi-threaded attack. The attack makes use of specific assembly code and jump addresses that needs to be set post compilation into the attack code.
The exact steps to achieve this are enumerated in the README file provided.
The attack can be run on same core or cross core using \texttt{S} or \texttt{C} keywords respectively while running the multiThreadIC binary.
As with the dcache\_poc, the exact build and run steps are provided in the README under icache\_poc/multiThread/ directory.

\subsection{Evaluation and Expected Results}
The PoC evaluates the attack as described in the paper. The scripts included help recreate Figures \ref{fig:dcuload} and \ref{fig:poc_results}. Figure \ref{fig:dcuload} shows measurements observed in the presence and absence of the attacker (interference gadget). In Figure \ref{fig:poc_results}, we show the bit error rate vs. bit rate (throughput).
Figure \ref{fig:poc_results} is created by running the experiment multiple times for a set number of bits and then taking the average time spent on each run versus the number of bits correctly leaked.

\subsection{Experiment Customization}
\paragraph{D-Cache PoC}
The configurable parts of the D-Cache PoC are
specified in the README. Important parameters for the success of the PoC include the fixed delay used to offset the load B, as well as the iterations of the dependency chain of instructions used to issue load A. The interaction between load A and load B needs to be configured to have an ordering influenced by the secret-dependent contention.

\paragraph{I-Cache PoC}
The I-Cache PoC can be customized for the number bit-leaks and the variant of the attack we want to run. All the customizations are detailed in the provided README.

\bibliographystyle{ACM-Reference-Format}
\bibliography{references.tex}

\end{document}